\documentclass[preprint,review,12pt]{elsarticle}
\usepackage{ifpdf}
\usepackage{graphicx}
\usepackage{amssymb}
\usepackage{amsmath}

\begin{document}

\begin{frontmatter}

\title{On certain new exact solutions of a diffusive predator-prey system}

\author[idf]{R.A.~Kraenkel}
\author[bdu]{K.~Manikandan}
\author[bdu]{M.~Senthilvelan\corref{auth1}}
\ead{velan@cnld.bdu.ac.in}
\address[idf]{Instituto de F\'{\i}sica Te\'orica , Universidade Estadual Paulista, Rua Dr. Bento Teobaldo Ferraz 271, 01140-070 S\~ao Paulo, Brazil}
\address[bdu]{Centre for Nonlinear Dynamics, School of Physics, Bharathidasan University, Tiruchirappalli 620024, Tamil Nadu, India}
\cortext[auth1]{Corresponding author; Phone : +91 431 2407057; Fax : +91 431 2407093}

\begin{abstract}
We construct exact solutions for a system of two nonlinear partial differential equations describing the spatio-temporal dynamics of a predator-prey system where the prey per capita growth rate is subject to the Allee effect.  Using the $\big(\frac{G'}{G}\big)$ expansion method, we derive exact solutions to this model for two different wave speeds.  For each wave velocity we report three different forms of solutions.  We also discuss the biological relevance of the solutions obtained.
\end{abstract}


\end{frontmatter}

\section{Introduction}
In this paper we derive exact solutions for a diffusive predator-prey system \cite{pet},
\begin{eqnarray}
u_t &=& u_{xx}-\beta u+(1+\beta)u^2-u^3-uv \nonumber \\
v_t &=& v_{xx}+kuv-mv-\delta v^3
\label{qqq1}
\end{eqnarray}
where $k$, $\delta$, $m$ and $\beta$ are positive parameters. Subscripts $x$ and $t$ denote partial derivatives.  The equations are expressed in dimensionless variables, where scaling  of space and time have been introduced so to have the equation in a simple form. The biological meaning of the each term has been discussed in \cite{pet}, which we briefly review.  The model is of the predator-prey kind,  with $u$ and $v$ being  the densities of prey and predator.  Spatial redistribution of the population is governed by diffusive dynamics, with both species having the same diffusivities,  \cite{okubo}.  In absence of the predator, the temporal dynamics of the  prey density  is of the Allee type \cite{all,cour},  small populations  being not viable. Presence of the predator population negatively affects the prey population. In turn, the predator population is totally dependent on prey availability,  as the only term in the predator equation that represents  the growth of the population is the $kuv$ term, where $k$ quantifies the gain in natality due to prey consumption.  The parameters $\beta$ and $m$ represent the per capita mortality rate of prey and predator respectively, in the linear, small populations, limit.  Finally, the  $\delta v^3$ is a closure relation taking into account the effects of higher trophic levels \cite{petr,steele}.

To investigate the dynamics of the above diffusive predator-prey system the authors of Ref. \cite{pet} have assumed the following relations between the parameters, namely $m=\beta$ and $k+\frac{1}{\sqrt{\delta}}=\beta+1$.  In other words it has been assumed that the per capita mortality rate of prey and predator are equal and the rate of biomass production at the lower level must be consistent with the rate of biomass assimilation at the upper level of the web \cite{dub, she, owe}.  Under this assumption Eq. (\ref{qqq1}) reads 
\begin{eqnarray}
\label{qqq6}
u_t &=& u_{xx}-\beta u+(k+\frac{1}{\sqrt{\delta}})u^2-u^3-uv \nonumber \\
v_t &=& v_{xx}+kuv-\beta v-\delta v^3.
\end{eqnarray}
In our work also we consider the predator-prey system (\ref{qqq6}) only. We construct exact analytic solutions to the equation (\ref{qqq6}) in order to understand the properties of this model for different parametric values.  To do so we employ the $\big(\frac{G'}{G}\big)$ expansion method \cite{bek,dai,wan,zay,zha}.  This method has been applied to several nonlinear evolutionary equations.  Here we demonstrate that the utility of this method in exploring dynamics of a diffusive predator-prey system.  While implementing  $\big(\frac{G'}{G}\big)$ expansion method to equation (\ref{qqq6}) we obtain exact solutions for two different wave speeds (vide Eqs. (\ref{qqq17}) and (\ref{qqq18})).  In both the cases we give three different types of exact solutions.

The plan of the paper is as follows.  To begin with, in Sec. 2, we describe the $\big(\frac{G'}{G}\big)$ expansion method.  In Sec. 3, we consider Eq. (\ref{qqq6}) and derive exact solutions of it.  Finally, we present our conclusion in Sec. 4.

\section{The $(\frac{G^{\prime}}{G})$-expansion method }
\label{sec:1}
In this section we discuss briefly the method of finding exact solutions for a system of nonlinear partial differential equations (PDEs) using $(\frac{G^{\prime}}{G})$-expansion method.

Suppose that the system of nonlinear PDEs is of the form
\begin{eqnarray}
\label{aaa1}
P(u,v,u_t,u_x,v_x,u_{tt},u_{xt},u_{xx},...)=0,\nonumber\\
Q(u,v,v_t,u_x,v_x,u_{tt},u_{xt},u_{xx},...)=0,
\end{eqnarray}
where $u=u(x,t)$ and $v=v(x,t)$ are two unknown functions and $P$ and $Q$ are polynomials in $u=u(x,t)$ and 
$v=v(x,t)$ and their partial
derivatives.  The $(\frac{G^{\prime}}{G})$ expansion method involves the following four steps.

${\bf Step\;1:}$ Let us introduce the travelling wave reduction
\begin{equation}
\label{aaa2}
u(x,t)=u(\xi), \;\; v(x,t)=v(\xi), \qquad \xi=x-ct,
\end{equation}
in the PDE (\ref{aaa1}) so that the latter becomes
\begin{eqnarray}
\label{aaa3}
P(u,v,-cu',u',v',c^2u'',-cu'',u'',...)=0,\nonumber\\
Q(u,v,-cv',u',v',c^2u'',-cu'',u'',...)=0,
\end{eqnarray}
where prime denotes differentiation with respect to the new variable $\xi$.

${\bf Step\;2:}$ Suppose that the solution of $(\ref{aaa3})$ can be expressed by a polynomial in $(\frac{G'}{G})$, that is
\begin{eqnarray}
\label{aaa4}
u(\xi)&=& \alpha_m\Big(\frac{G'}{G}\Big)^m+\alpha_{m-1}\Big(\frac{G'}{G}\Big)^{m-1}
+\alpha_{m-2}\Big(\frac{G'}{G}\Big)^{m-2}+...,\nonumber\\
v(\xi)&=& \beta_n\Big(\frac{G'}{G}\Big)^n+\beta_{n-1}\Big(\frac{G'}{G}\Big)^{n-1}
+\beta_{n-2}\Big(\frac{G'}{G}\Big)^{n-2}+...,
\end{eqnarray}
with $G=G(\xi)$ is the solution of the second order linear damped harmonic oscillator equation
\begin{equation}
\label{aaa5}
G''+\lambda G'+\mu G=0.
\end{equation}
In the above $\alpha_m, \beta_n, \alpha_{m-1}, \beta_{n-1}..., \alpha_0, \beta_0, \lambda,$ and $\mu$ are constants 
and $\alpha_m \neq 0$, $\beta_n \neq 0$.  The positive integers $m$ and $n$ can be determined by substituting (\ref{aaa4}) in (\ref{aaa3}) and considering the homogeneous
balance between the highest order derivative and nonlinear terms appearing in (\ref{aaa3}).

${\bf Step\;3:}$ Substituting (\ref{aaa4}) in (\ref{aaa3}) and eliminating the variable $G''$ in the resultant equations by using (\ref{aaa5}) one gets two polynomials equations in $(\frac{G'}{G})$.  Now equating each coefficients of $(\frac{G'}{G})^m$ and $(\frac{G'}{G})^n$ to zero one obtains a set of algebraic equations for the parameters $\alpha_m, \beta_n, \alpha_{m-1}, \beta_{n-1}..., \alpha_0, \beta_0, \lambda,$ and $\mu$.  Solving these algebraic equations one can get exact values for these coefficients.

${\bf Step\;4:}$ Substituting the values of $\alpha_m, \beta_n, \alpha_{m-1}, \beta_{n-1}..., \alpha_0, \beta_0, \lambda,$ and $\mu$ and $c$ and the general solution of (\ref{aaa5}) in (\ref{aaa4}) one can obtain three different types of travelling wave solutions for the given system of nonlinear PDEs.

\section {Diffusive predator-prey system}
In this section, we apply the method described in the previous section to the nonlinear PDEs (\ref{qqq6})
and construct exact solutions.  Substituting (\ref{aaa2}) into  (\ref{qqq6}) we get the following system of ordinary differential equations (ODEs), namely
\begin{eqnarray}
\label{qqq9}
u''+cu'-\beta u+\big(k+\frac{1}{\sqrt{\delta}}\big) u^2-u^3-uv=0, \nonumber \\
v''+cv'+kuv-\beta v-\delta v^3=0.
\end{eqnarray}
Suppose that the solution of ODEs (\ref{qqq9}) can be expressed by a polynomial in $(\frac{G'}{G})$ which is of the form (\ref{aaa4}).

Substituting (\ref{aaa4}) and their derivatives in (\ref{qqq9}) and performing the homogeneous balance between $u''$ and $u^3$ and $v''$ with $v^3$ in resultant equation we find $m=1$ and $n=1$.  So we fix the polynomials (\ref{aaa4}) be of the form
\begin{eqnarray}
\label{qqq13}
u(\xi)=\alpha_1 \Big(\frac{G'}{G}\Big)+\alpha_0, \;\;
v(\xi)=\beta_1 \Big(\frac{G'}{G}\Big)+\beta_0, \;\; \alpha_1, \beta_1 \neq 0.
\end{eqnarray}
Substituting the expressions (\ref{qqq13}) and their derivatives in (\ref{qqq9}) and rearranging the resultant equation in the descending powers of $\Big(\frac{G'}{G}\Big)$ we arrive at
\begin{align}
\label{qqq15}
 [2\alpha_1-\alpha_1^3]\Big(\frac{G'}{G}\Big)^3+[3\alpha_1\lambda-c\alpha_1+k\alpha_1^2+\frac{\alpha_1^2}{\sqrt{\delta}}-3\alpha_1^2\alpha_0-\alpha_1\beta_1]\Big(\frac{G'}{G}\Big)^2 && \nonumber \\ 
 +[(2\mu+\lambda^2)\alpha_1-c\lambda \alpha_1-\beta\alpha_1+2k\alpha_0\alpha_1+\frac{2\alpha_1\alpha_0}{\sqrt{\delta}}-3\alpha_0^2\alpha_1-\alpha_1\beta_0-\alpha_0\beta_1]\Big(\frac{G'}{G}\Big) && \nonumber \\ 
 +(\mu \alpha_1\lambda-c\mu \alpha_1-\beta \alpha_0+\alpha_0^2+\frac{\alpha_0^2}{\sqrt{\delta}}-\alpha_0^3-\alpha_0\beta_0) = 0, \qquad \qquad \qquad
\end{align}
\begin{eqnarray}
\label{qqq16}
 [2\beta_1-\delta\beta_1^3]\Big(\frac{G'}{G}\Big)^3+[3\beta_1\lambda-c\beta_1+k\alpha_1\beta_1-3\delta\beta_1^2\beta_0]\Big(\frac{G'}{G}\Big)^2 && \nonumber \\ 
 +[(2\mu+\lambda^2)\beta_1-c\lambda \beta_1-\beta\beta_1+k\alpha_0\beta_1+k\alpha_1\beta_0-3\delta\beta_0^2\beta_1]\Big(\frac{G'}{G}\Big) && \nonumber \\ 
 +(\mu \beta_1\lambda-c\mu \beta_1-\beta \beta_0+k\alpha_0\beta_0-\delta\beta_0^3) = 0. \qquad  \qquad  
\end{eqnarray}
Equating the coefficients of $(\frac{G'}{G})^m$, $m=0,1,2,3,$ to zero in equations (\ref{qqq15}) and (\ref{qqq16}) we get the following set of algebraic equations, namely
\begin{align}
\label{qqa17}
& 2\alpha_1-\alpha_1^3 =0, \nonumber \\ 
&3\alpha_1\lambda-c\alpha_1+k\alpha_1^2+\frac{\alpha_1^2}{\sqrt{\delta}}-3\alpha_1^2\alpha_0-\alpha_1\beta_1=0, \nonumber \\ 
&(2\mu+\lambda^2)\alpha_1-c\lambda \alpha_1-\beta\alpha_1+2k\alpha_0\alpha_1+\frac{2\alpha_1\alpha_0}{\sqrt{\delta}}-3\alpha_0^2\alpha_1-\alpha_1\beta_0-\alpha_0\beta_1 =0, \nonumber \\ 
&\mu \alpha_1\lambda-c\mu \alpha_1-\beta \alpha_0+\alpha_0^2+\frac{\alpha_0^2}{\sqrt{\delta}}-\alpha_0^3-\alpha_0\beta_0 =0.\\
\label{qqa18}
&2\beta_1-\delta\beta_1^3 =0,  \nonumber \\ 
&3\beta_1\lambda-c\beta_1+k\alpha_1\beta_1-3\delta\beta_1^2\beta_0 =0, \nonumber \\ 
&(2\mu+\lambda^2)\beta_1-c\lambda \beta_1-\beta\beta_1+k\alpha_0\beta_1+k\alpha_1\beta_0-3\delta\beta_0^2\beta_1 =0, \nonumber \\ 
&\mu \beta_1\lambda-c\mu \beta_1-\beta \beta_0+k\alpha_0\beta_0-\delta\beta_0^3 =0. 
\end{align}
Solving the above system of algebraic equations (\ref{qqa17}) and (\ref{qqa18}) we obtain two sets of values for the constants $\alpha_1$, $\alpha_0$, $\beta_1$, $\beta_0$ and $c$:
\begin{eqnarray}
\label{qqq17}
(a)\quad  \alpha_1=\pm \sqrt{2}, \; \beta_0=\frac{\alpha_0}{\sqrt{\delta}}, \; \beta_1=\pm \sqrt{\frac{2}{\delta}}, \; c=\mp \frac{k}{\sqrt{2}},\; \lambda=\mp \frac{k-2\alpha_0}{\sqrt{2}},\nonumber \\  \beta=k\alpha_0-\alpha_0^2+2\mu. \qquad \qquad \qquad \qquad \qquad \qquad \qquad \qquad \qquad
\end{eqnarray}
\begin{eqnarray}
\label{qqq18}
(b)\quad \alpha_1=\pm \sqrt{2},\; \beta_0=\frac{\alpha_0}{\sqrt{\delta}}, \; \beta_1=\pm \sqrt{\frac{2}{\delta}}, \; \lambda=\pm \frac{\alpha_0^2+2\mu}{\sqrt{2}\alpha_0},\qquad \qquad \quad \nonumber \\ \quad c=\pm \frac{1}{\sqrt{2}}\big(2k-3\alpha_0+\frac{6\mu}{\alpha_0}\big),\; \;\beta=-\frac{(\alpha_0^2-2\mu)(-k\alpha_0+\alpha_0^2-2\mu)}{\alpha_0^2}.  
\end{eqnarray}
Since both the sets separately satisfy the algebraic equations in $(\ref{qqa17})$ and $(\ref{qqa18})$ they individually form a compatible solution.  From each set we derive an exact solution for the nonlinear PDEs (\ref{qqq6}).

To begin with let us take the values given in $(\ref{qqq17}).$
With the values given in $(\ref{qqq17})$, the solution $(\ref{qqq13})$ reads 
\begin{eqnarray}
\label{qqq19}
u(\xi)=\pm \sqrt{2}\bigg(\frac{G'}{G}\bigg)+\alpha_0, \;\;\;
v(\xi)=\pm \sqrt{\frac{2}{\delta}}\bigg(\frac{G'}{G}\bigg)+\frac{\alpha_0}{\sqrt{\delta}}.
\end{eqnarray}
It is known that the linear damped harmonic oscillator equation (\ref{aaa5}) admits three different types of solutions depending on the
values of $\lambda$ and $\mu$, namely

Case 1: $\;\lambda^2-4\mu > 0$
\begin{equation}
\label{qa17}
G(\xi) = e^{(-\lambda/2)\xi}\Big(c_1\sinh \frac{\sqrt{\lambda^2-4\mu}}{2}\xi+ c_2\cosh \frac{\sqrt{\lambda^2-4\mu}}{2}\xi\Big)
\end{equation}
Case 2:$\;\lambda^2-4\mu < 0$
\begin{equation}
\label{qa18}
G(\xi) = e^{(-\lambda/2)\xi}\Big(c_1\cos \frac{\sqrt{4\mu-\lambda^2}}{2}\xi+ c_2\sin \frac{\sqrt{4\mu-\lambda^2}}{2}\xi\Big)
\end{equation}
Case 3: $\lambda^2-4\mu = 0$
\begin{equation}
\label{qa19}
G(\xi) = (c_1+c_2\xi)e^{(-\lambda/2)\xi}.\qquad \qquad \qquad \qquad \qquad \qquad \quad
\end{equation}

Substituting (\ref{qa17})-(\ref{qa19}) into (\ref{qqq19}) we arrive at the following form of solutions, namely

Case 1:  $\lambda^2-4\mu \;\textgreater\; 0 $
\begin{align}
\label{qqq21}
\quad u(\xi)=\pm \sqrt{2}\Bigg(-\frac{\lambda}{2}+\frac{\sqrt{\lambda^2-4\mu}}{2}\Bigg(\frac{c_1\cosh \frac{\sqrt{\lambda^2-4\mu}}{2}\xi+c_2\sinh \frac{\sqrt{\lambda^2-4\mu}}{2}\xi}{c_1\sinh \frac{\sqrt{\lambda^2-4\mu}}{2}\xi+ c_2\cosh \frac{\sqrt{\lambda^2-4\mu}}{2}\xi}\Bigg)\Bigg)+\alpha_0,\nonumber \\
\quad v(\xi)=\pm \sqrt{\frac{2}{\delta}}\Bigg(-\frac{\lambda}{2}+\frac{\sqrt{\lambda^2-4\mu}}{2}\Bigg(\frac{c_1\cosh \frac{\sqrt{\lambda^2-4\mu}}{2}\xi+c_2\sinh \frac{\sqrt{\lambda^2-4\mu}}{2}\xi}{c_1\sinh \frac{\sqrt{\lambda^2-4\mu}}{2}\xi+ c_2\cosh \frac{\sqrt{\lambda^2-4\mu}}{2}\xi}\Bigg)\Bigg)+\frac{\alpha_0}{\sqrt{\delta}}.
\end{align}
\begin{figure}
\begin{center}
\includegraphics[width=.70\linewidth]{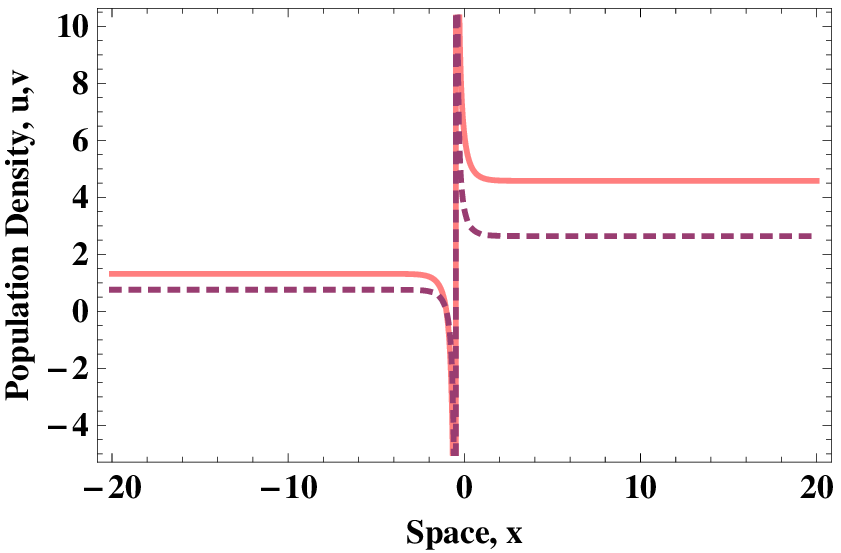}
\end{center}
\caption{The densities of prey (solid line) and predator (dashed line)  as given by the exact solution (\ref{qqq21}) shown for time $t=0$; when $\lambda\textgreater2\sqrt{\mu}$; Parameters are $\alpha_0=1.2$, $k=5.9$, $\delta=3$, $\mu=0.2$, $c_1=20$, $c_2=10$.}
\end{figure}
\begin{figure}
\begin{center}
\includegraphics[width=.70\linewidth]{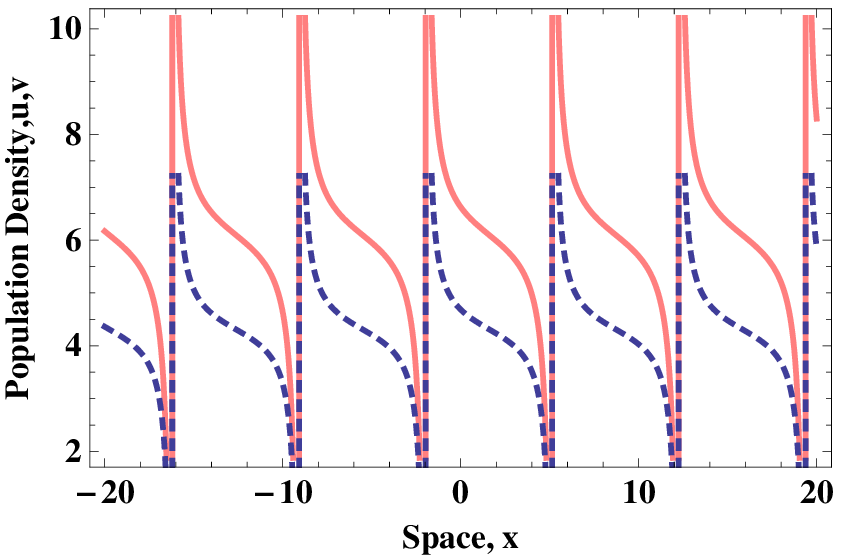}
\end{center}
\caption{The densities of prey (solid line) and predator (dashed line)  as given by the exact solution (\ref{qqq23}) shown for time $t=50$; when $\lambda\textless2\sqrt{\mu}$; Parameters are $\alpha_0=3$, $k=12.2$, $\delta=2$, $\mu=5$, $c_1=20$, $c_2=-10$.}
\label{s2}
\end{figure}
\begin{figure}
\begin{center}
\includegraphics[width=.65\linewidth]{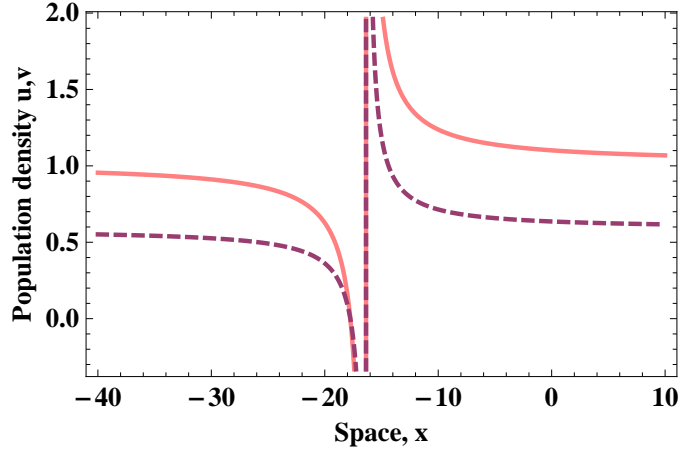}
\end{center}
\caption{The densities of prey (solid line) and predator (dashed line)  as given by the exact solution (\ref{qqq25}) shown for time $t=10$
; when $\lambda=2\sqrt{\mu}$; Parameters are $\delta=3$, $k=2.03$, $\mu=1$, $c_1=20$, $c_2=10$.}
\end{figure}

Case 2: $\lambda^2-4\mu \;\textless\; 0 $
\begin{align}
\label{qqq23}
\quad u(\xi)=\pm \sqrt{2}\Bigg(-\frac{\lambda}{2}+\frac{\sqrt{4\mu-\lambda^2}}{2}\Bigg(\frac{-c_1\sin \frac{\sqrt{4\mu-\lambda^2}}{2}\xi+c_2\cos \frac{\sqrt{4\mu-\lambda^2}}{2}\xi}{c_1\cos \frac{\sqrt{4\mu-\lambda^2}}{2}\xi+ c_2\sin \frac{\sqrt{4\mu-\lambda^2}}{2}\xi}\Bigg)\Bigg)+\alpha_0, \nonumber \\
\quad v(\xi)=\pm \sqrt{\frac{2}{\delta}}\Bigg(-\frac{\lambda}{2}+\frac{\sqrt{4\mu-\lambda^2}}{2}\Bigg(\frac{-c_1\sin \frac{\sqrt{4\mu-\lambda^2}}{2}\xi+c_2\cos \frac{\sqrt{4\mu-\lambda^2}}{2}\xi}{c_1\cos \frac{\sqrt{4\mu-\lambda^2}}{2}\xi+ c_2\sin \frac{\sqrt{4\mu-\lambda^2}}{2}\xi}\Bigg)\Bigg)+\frac{\alpha_0}{\sqrt{\delta}}.
\end{align}

Case 3: $\lambda^2-4\mu=0$
\begin{eqnarray}
\label{qqq25}
\quad u(\xi)=\pm \sqrt{2}\Bigg(\frac{c_2}{c_1+c_2\xi}-\frac{\lambda}{2}\Bigg)+\alpha_0, \qquad \qquad \qquad \qquad \qquad \qquad \qquad \qquad \nonumber \\
\quad v(\xi)=\pm \sqrt{\frac{2}{\delta}}\Bigg(\frac{c_2}{c_1+c_2\xi}-\frac{\lambda}{2}\Bigg)+\frac{\alpha_0}{\sqrt{\delta}}.\qquad \qquad \qquad \qquad \qquad \qquad \qquad \quad
\end{eqnarray}
where $\alpha_0=-\frac{2\sqrt{2\mu}+k}{2}$, 
$\xi=x\pm\Big(\frac{k}{\sqrt{2}}\Big)t $, $c_1$ and $c_2$ are arbitrary constants.

For the second set of values we end up with the same form of solution with the only difference in the values in the parameters $c$, $\beta$ and $\lambda$:

Case 1:  $\lambda^2-4\mu \; \textgreater\; 0 $
\begin{align}
\label{qqq22}
\quad u(\xi)=\pm \sqrt{2}\Bigg(-\frac{\lambda}{2}+\frac{\sqrt{\lambda^2-4\mu}}{2}\Bigg(\frac{c_1\cosh \frac{\sqrt{\lambda^2-4\mu}}{2}\xi+c_2\sinh \frac{\sqrt{\lambda^2-4\mu}}{2}\xi}{c_1\sinh \frac{\sqrt{\lambda^2-4\mu}}{2}\xi+ c_2\cosh \frac{\sqrt{\lambda^2-4\mu}}{2}\xi}\Bigg)\Bigg)+\alpha_0,\nonumber \\
\quad v(\xi)=\pm \sqrt{\frac{2}{\delta}}\Bigg(-\frac{\lambda}{2}+\frac{\sqrt{\lambda^2-4\mu}}{2}\Bigg(\frac{c_1\cosh \frac{\sqrt{\lambda^2-4\mu}}{2}\xi+c_2\sinh \frac{\sqrt{\lambda^2-4\mu}}{2}\xi}{c_1\sinh \frac{\sqrt{\lambda^2-4\mu}}{2}\xi+ c_2\cosh \frac{\sqrt{\lambda^2-4\mu}}{2}\xi}\Bigg)\Bigg)+\frac{\alpha_0}{\sqrt{\delta}}.
\end{align}
Case 2: $\lambda^2-4\mu \;\textless\;0 $
\begin{align}
\label{qqq24}
\quad u(\xi)=\pm \sqrt{2}\Bigg(-\frac{\lambda}{2}+\frac{\sqrt{4\mu-\lambda^2}}{2}\Bigg(\frac{-c_1\sin \frac{\sqrt{4\mu-\lambda^2}}{2}\xi+c_2\cos \frac{\sqrt{4\mu-\lambda^2}}{2}\xi}{c_1\cos \frac{\sqrt{4\mu-\lambda^2}}{2}\xi+ c_2\sin \frac{\sqrt{4\mu-\lambda^2}}{2}\xi}\Bigg)\Bigg)+\alpha_0, \nonumber \\
\quad v(\xi)=\pm \sqrt{\frac{2}{\delta}}\Bigg(-\frac{\lambda}{2}+\frac{\sqrt{4\mu-\lambda^2}}{2}\Bigg(\frac{-c_1\sin \frac{\sqrt{4\mu-\lambda^2}}{2}\xi+c_2\cos \frac{\sqrt{4\mu-\lambda^2}}{2}\xi}{c_1\cos \frac{\sqrt{4\mu-\lambda^2}}{2}\xi+ c_2\sin \frac{\sqrt{4\mu-\lambda^2}}{2}\xi}\Bigg)\Bigg)+\frac{\alpha_0}{\sqrt{\delta}}.
\end{align}
Case 3: $\lambda^2-4\mu=0$
\begin{eqnarray}
\label{qqq26}
\quad u(\xi)=\pm \sqrt{2}\Bigg(\frac{c_2}{c_1+c_2\xi}-\frac{\lambda}{2}\Bigg)+\alpha_0, \qquad \qquad \qquad \qquad \qquad \qquad \qquad \qquad \nonumber \\
\quad v(\xi)=\pm \sqrt{\frac{2}{\delta}}\Bigg(\frac{c_2}{c_1+c_2\xi}-\frac{\lambda}{2}\Bigg)+\frac{\alpha_0}{\sqrt{\delta}}. \qquad \qquad \qquad \qquad \qquad \qquad \qquad \quad
\end{eqnarray}
where $\alpha_0=\sqrt{2\mu}$, 
$\xi=x\mp \frac{1}{2}\Big(2\sqrt{2}k-3\sqrt{2}\alpha_0+\frac{6\sqrt{2}\mu}{\alpha_0}\Big)$, $c_1$ and $c_2$ are arbitrary constants.
{\bf \section {Discussion and Conclusion}}
In this paper, we have constructed exact solutions for a diffusive predator-prey system which is modeled by a system of two coupled nonlinear PDEs.  Using  the $\big(\frac{G'}{G}\big)$ expansion method, we have derived exact solutions for two different wave speeds.  

The solutions that we have obtained are singular and cannot be taken at face value describing actual situations in ecology. Notwithstanding they present a very interesting property: depending on the sign of $\lambda^2 -4\mu$ , the nature of the solution changes from a single structure to a periodic one, which is akin to pattern forming systems.   

We express $\lambda^2 -4\mu$ in terms of the original parameters  in the equation, we get $\lambda^2 -4\mu = k/2-2\beta$.  Therefore, if $k^2>4\beta$ we have a single structure like in Fig.(1). And if  $k^2<4\beta$ we have a periodic structure. The meaning of $k$ is that it measures the gain in natality obtained by the predator, and $\beta$ is its mortality.  It follows that periodic structure formation comes from the strength  of mortality. In a more intuitive way,  $\beta $ is the inverse of the typical time for a predator population to decay in absence of preys. If this time is short,  we have a periodic pattern, the population continuously decaying and recovering. If this time is long, a smoother dynamics shows up.

The equations for which we could find new solutions, besides the previously known, \cite{pet}, have the obvious  drawback that matching of coefficients is necessary. This is the same situation in most cases in the subject of exact solutions for reaction-diffusion equations, beyond the specific cases of interest in biology, \cite{we}. However, the broad view insight gained remains of interest as the systems considered contain many elements of more realistic, non solvable ones.

{\bf \section*{Acknowledgements}}
RAK and MS wish to thank CNPq (Brazil) and DST (India) for the financial support through major research projects.

\label{lastpage}

\end{document}